\documentclass{ws-p8-50x6-00}
\def\be {\begin{equation}}
\def\ee {\end{equation}}
\def\beqy{\begin{eqnarray}}
\def\eeqy{\end{eqnarray}}

\def\jpsi{J/\psi}
\def\etac{\eta_c}
\def\as{\alpha_s}

\begin{document}

\title{Two Photon Width of $\eta_c$}

\author{Nicola Fabiano}
\address{ Perugia University and INFN, via Pascoli I-06100, Perugia, Italy}
\author{Giulia Pancheri}
\address{INFN National Laboratories, P.O.Box 13, I00044 Frascati, Italy}


\maketitle

\abstracts{We discuss the measured partial width of the pseudoscalar
charmonium state, $\eta_c$ , into two photons.
Predictions from potential models are examined and compared with
experimental values. Including radiative corrections, it is found
that present measurements are compatible both with a 
QCD type potential and with a static Coulomb potential, with
$\alpha_s$ evaluated at two loops. Results are also compared with those
from $\jpsi$ data through the NRQCD model.}

\section{Introduction}\label{subsec:intro}
 In this note,
we examine the theoretical predictions for the electromagnetic
decay of the simplest
and lowest lying of all the charmonium states, i.e. the pseudoscalar 
$\eta_c$~.  We shall compare the two photon decay width with 
 the leptonic width of the $J/\psi$, which has been  
measured with  higher precision~\cite{PSI} and found to be 15\%
 higher than in previous measurements \cite{PDGPSI}.
The most recently  reported  Particle Data Group average \cite{PDG}
is given by
\be
\Gamma_{exp}(\eta_c \to \gamma \gamma)=7.4 \pm 1.4  \textrm{ keV}
\label{eq:etactogamgam}
\ee
\section{Relation to $J/\psi \to e^+e^-$ width}\label{subsec:otherwidth}
The two photon decay width of a pseudoscalar
quark-antiquark bound state can be written as~\cite{VANROYEN,BARBIERI}
\be
 \Gamma(\eta_c\rightarrow \gamma\gamma)=12
e_q^4\alpha_{QED}^2
4\pi \frac{|\psi(0)|^2}{M^2} \left [ 1 + \frac{\alpha_s}{\pi} 
\left(\frac{\pi^2-20}{3} \right ) \right ] \approx\Gamma_B^P \left 
(1 -\alpha_s \right )\label{eq:widsc}
\ee
where $\psi(0)$ is the wavefunction of the interquark potential
evaluated at the origin.
It is useful to compare eq.~(\ref{eq:widsc})  with the expressions for the 
vector state $J/\psi$~\cite{MACKENZIE}, i.e.
\be
\Gamma(J/\psi\rightarrow ee)=4 
e_q^2\alpha_{QED}^2
4\pi \frac{|\psi(0)|^2}{M^2}\left ( 1-
\frac{16}{3}\frac{\alpha_s}{\pi} \right )
\approx \Gamma_B^V \left ( 1-1.7 \alpha_s \right ) .
\label{eq:widvec}
\ee
The expressions in eq. (\ref{eq:widsc}) and (\ref{eq:widvec}) can be 
used to estimate 
the radiative width of $\eta_c$ from the measured values of the
leptonic decay width of $J/\psi$, if one assumes that the $\psi(0)$
values for both the pseudoscalar and the vector state 
should be the same. This is true up to errors of $\mathcal{O}(\alpha_s/m_c^2)$.
From
\be
\Gamma_{exp} (J/\psi\rightarrow e^+e^-)= 5.26 \pm 0.37 \textrm{ keV}
\label{eq:expjpsiee}
\ee
expanding in $\as$ one has:
\be
\frac{\Gamma (\eta_c\rightarrow \gamma\gamma)}{\Gamma(J/\psi
\rightarrow e^+e^-)}\approx
\frac{4}{3} \frac{(1-3.38\as/\pi)}{(1-5.34\as/\pi)} = 
\frac{4}{3}  \left [ 1+1.96  \frac{\alpha_s}{\pi} 
+ \mathcal{O}(\alpha_s^2) \right ]
\label{eq:rappwid}
\ee
From the value $\alpha_s(M_Z) = 0.118\pm 0.003 $
the renormalization group evolution gives
$\alpha_s(Q=2m_c=3.0\mbox{ GeV})=0.25 \pm 0.01$.
Combining the formul\ae{}~(\ref{eq:expjpsiee}) and~(\ref{eq:rappwid})
we obtain
\be
\Gamma(\eta_c\rightarrow \gamma\gamma) \pm
\Delta\Gamma(\eta_c\rightarrow \gamma\gamma) =8.18
\underbrace{\pm 0.57}_{\scriptstyle J/\psi \textrm{\small{ error}} }  
\overbrace{\pm 0.04}^{\scriptstyle \as \textrm{\small{ error}}} \ \ \textrm{keV}
\ee
This estimate agrees within $1 \sigma$ with the value given
in formula~(\ref{eq:etactogamgam}).
\section{Potential models predictions}
We shall extract now the wave function at the origin from potential models.
For the calculation of the wavefunction we have used four different 
potential models,
like the Cornell type potential \cite{CORNELL}
$V(r) = -\frac{k}{r} + \frac{r}{a^{2}} $
with parameters $a=2.43$, $k=0.52$, the Richardson potential 
\cite{RICHARDSON}
$V_{R}(r) = -\frac{4}{3} \frac{12 \pi}{33-2N_{f}} \int 
\frac{d^{3}q}{(2\pi)^{3}} \frac{e^{iqr}}{q^{2}\log(1+q^{2}/
\Lambda^{2})}$
with $N_{f} = 3$ , $ \Lambda=398$ $MeV$, and 
the QCD inspired potential $V_{J}$  of Igi-Ono \cite{IGI,TYE}
$ V_{J}(r) = V_{AR}(r) + d r e^{-gr} + ar , \ 
 V _{AR}(r) = -\frac{4}{3} \frac{\alpha_{s}^{(2)}(r)}{r} $ 
with two different  parameter sets, corresponding to 
 $\Lambda_{\overline{MS}}=0.5 \  GeV$
and $\Lambda_{\overline{MS}}=0.3 \  GeV$ respectively \cite{IGI}. We also 
show the results from a 
Coulombic type potential with the QCD coupling $\alpha_s$ frozen
to a value of $r$ corresponding to the Bohr radius of the
quarkonium system, $r_B=3/(2m_c\as)$ (see for instance \cite{NOI}). 
The error sources in calculation are given by the
choice of scale in radiative correction,
the choice of various potential parameters
and the fluctuations in results from different models.
The $\Gamma(\etac \to \gamma \gamma)$ potential models prediction gives a range
of values:
\be
\Gamma(\etac \to \gamma \gamma) =7.6 \pm 1.5 \textrm{ keV} 
\label{eq:potmodpred}
\ee

\section{Octet Component model}

We will  present now another model which admits other components to the meson
decay beyond the one from the colour singlet picture (Bodwin, Braaten and
Lepage)~\cite{BBL}. 
NRQCD has been used to separate the short distance scale of 
annihilation from the nonperturbative contributions of long distance scale.
This model has been successfully used to
explain the larger than expected $\jpsi$ production at the
Tevatron.
According to~\cite{BBL}, in the octet model for quarkonium the 
decay widths of charmonium states involve four unknown long distance 
coefficients which can be reduced to
two by means of the vacuum saturation approximation:
$G_1 \equiv \langle \jpsi | O_1(^3S_1) | \jpsi \rangle = \langle \etac 
| O_1(^1S_0) | \etac \rangle$ and
$F_1 \equiv \langle \jpsi | P_1(^3S_1) | \jpsi \rangle = \langle \etac 
| P_1(^1S_0) | \etac \rangle$,
correct up to $\mathcal{O}(v^2)$, the velocity of the quarks inside the meson.
We use the $J/\psi$ experimental decay widths as input in order to determine
the long distance coefficients $G_1$ and $F_1$~. This result in turn
is used to compute the $\etac$ decay widths.

The BBL model gives the following decay width of the $\etac$
meson:
\be
\Gamma(\eta_c\rightarrow \gamma\gamma)  =9.02
\underbrace{\pm 0.65}_{\scriptstyle J/\psi \textrm{\small{ error}} }  
\overbrace{\pm 0.14}^
{\scriptstyle \as \textrm{\small{ error}}} \textrm{ keV}
\ee
This value agrees with experimental data within $1\sigma$.
\section{Summary}

\begin{figure}[h]
\epsfxsize=15pc 
\begin{center}
\epsfbox{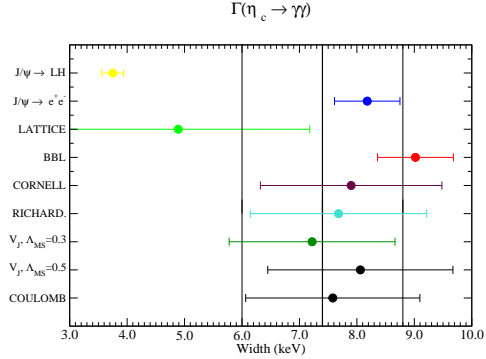} 
\end{center}
\caption{Potential Models results; BBL model with input from $J/\psi$ 
decay data; Lattice evaluation of
$G_1$ and $F_1$ factors; Singlet picture with $G_1$ obtained from 
$J/\psi \to e^+e^-$ and $J/\psi \to LH$ processes respectively. The vertical
lines represent the PDG average value and its indetermination.
  \label{fig:summary}}
\end{figure}

We present in fig.~(\ref{fig:summary})
a set of predictions coming from different
methods: results from potential models;  BBL model with $G_1$
and $F_1$ extracted from the $J/\psi$ decay data; lattice calculation of 
the long distance terms for the BBL model~\cite{LATTICE};
singlet picture: $G_1$ extracted from $J/\psi \to e^+e^-$ decay width, and
singlet picture: $G_1$ extracted from $J/\psi \to LH$ decay width.
\section{Conclusions}
The $\Gamma(\etac \to \gamma \gamma)$ decay width prediction of the 
potential models considered gives the value
 $7.6 \pm 1.5 \textrm{ keV}$
which is consistent with the PDG average. The Coulombic model is in agreement
with other models prediction.
Predictions of the BBL model for the $\etac \to \gamma \gamma$
decay width is consistent with the experimental measures, for both the long
distance terms $G_1$ and $F_1$ extracted from the $J/\psi$ experimental decay
widths and the one evaluated from lattice calculations.

\end{document}